\newcommand{\ltae}{\raisebox{-0.6ex}{$\,\stackrel
{\raisebox{-.2ex}{$\textstyle <$}}{\sim}\,$}}
\newcommand{\gtae}{\raisebox{-0.6ex}{$\,\stackrel
{\raisebox{-.2ex}{$\textstyle >$}}{\sim}\,$}}

\documentstyle{l-aa}
\begin{document}
\thesaurus{ 02(11.01.2, 13.18.1)}

\title{\Large\bf Radiative ages in a representative sample of low luminosity
radio galaxies}

\author{Parma P.\inst{1}, Murgia M. \inst{1,5}, Morganti R.  \inst{1}, 
Capetti A.\inst{2}, de Ruiter H.R.\inst{1,3}, Fanti R.\inst{1,4} }

\institute{Istituto di Radioastronomia, CNR, via Gobetti 101, I-40129
Bologna, Italy
\and Osservatorio Astronomico di Torino, Strada Osservatorio 20,
I-10025 Pino Torinese, Italy
\and Osservatorio Astronomico, via Zamboni 33, I-40126   Bologna, Italy
\and Dipartimento di Fisica, Universit\`a di Bologna, via Irnerio 46, I-40126 Bologna, Italy
\and  Dipartimento di Astronomia, Universit\`a di Bologna, Via Zamboni 33, I-40126 Bologna, Italy
 }

\maketitle
\markboth{Radiative ages of low luminosity radio galaxies}{Parma et al.}

\offprints{P. Parma}



\vskip 2 cm

\begin{abstract}
Two frequency observations, mainly at 1.4 and 5 GHz from the VLA, have been
used to study spectral variations along the lobes of some nearby low luminosity radio 
galaxies that constitute a representative sample selected from the B2 catalogue.
The variations of the spectral index have been interpreted as being due to 
synchrotron and inverse Compton losses and characteristic spectral ages 
are deduced for the relativistic electrons.
The $\it radiative ~ ages$  are in the range of several $10^7$ years.
These $\it ages $ correlate well with the source sizes.
They also appear to be consistent with $\it dynamical ~ ages$ determined from ram-pressure
arguments, if we make reasonable assumptions about the
ambient gas density and allow for very moderate deviations from the
equipartition conditions.
There appears to be a significant difference between the radiative ages
of sources in our sample and those of more powerful 3CR radio sources. 
We briefly discuss the possibility of re-acceleration processes and 
indicate some objects where these may occur.
\end{abstract}

\section{Introduction}

The determination of ages of extragalactic radio sources is one of the key
points for any theoretical model that wants to explain their origin and evolution.
As the radiation is due to the synchrotron process by relativistic electrons 
spiraling in magnetic field, much of the effort in determining source ages 
goes into the study of the radio spectra, for which the synchrotron 
theory predicts a frequency break due to the radiative energy losses, which 
drifts in time. According to various source evolution models, relativistic 
electrons in different regions of the source are deposited there at 
different times, so that the frequency break effectively is
a clock that indicates the time elapsed since their production.

For several years spectral studies were based on the integrated spectrum 
of the radio sources (see e.g. Kellerman 1964; van der Laan \& Perola 1969).
The advent of high resolution interferometric systems, like the Cambridge 5-km Telescope,
the Westerbork Synthesis Radio Telescope (WSRT) and the Very Large Array (VLA),
has made it possible to study the spectral behaviour in different regions of a source.
For a limited number of objects, detailed studies of the radio spectra across 
the emitting regions (see e.g. Alexander 1987; Carilli et al. 1991; 
Feretti  et al. 1998) have produced break frequency maps and, from them, 
source age maps.
However, in the majority of these studies the radio spectra across a source 
are based on only one pair of frequencies, so that the break frequency cannot
be seen directly from the data. Nevertheless, with some additional assumptions
which find their justification from the results obtained from the well
studied objects, the break frequency can be estimated with reasonable accuracy. 
In fact, the large body of data now available in literature is obtained this way
(see e.g. Alexander \& Leahy 1987; Leahy et al. 1989; Liu et al. 1992, 
for a large set of data on powerful 3CR sources).

Recently more sophisticated methods of analyzing spectral 
maps have been developed by Katz-Stone et al. (1993), Katz-Stone \& Rudnick
(1994) and Rudnick et al. (1994).  These authors point out that the traditional
method may lead to misinterpretation. 
In addition, Eilek \& Arendt (1996) consider synchrotron spectra arising
from a distribution of magnetic field strength and show  that high frequency 
steepening of the spectrum need not necessarily be due to synchrotron ageing.
Nevertheless we follow here the 
traditional spectral analysis, since this will facilitate comparison with the bulk 
of the literature in which this kind of analysis is very common. Moreover, the limited 
quality of our data does not allow anything more sophisticated.

Alternative methods for age determination are based on dynamical arguments, in particular
the balance between the jet thrust and the ram-pressure due to the external 
medium (see e.g. Carilli et al. 1991 for Cygnus A).
There is some debate about the consistency of the results between the two 
methods (see again Carilli et al. 1991 and Eilek 1996).

In this paper we present a two frequency spectral study of a 
representative subsample of the B2 radio galaxies and discuss
radiative ageing of the relativistic electrons caused by synchrotron
and inverse Compton (I.C.) energy losses. The present study is complementary 
those quoted above, which deal with powerful 3CR radio sources: the B2 radio sources are
less powerful by about two orders of magnitude\footnote{We use 
$\rm H_0 = 100$~km~s$^{-1}$Mpc$^{-1}$}.

The data on which this work is based and the 
spectral properties are described in Sect. 2 and 3.
In Sect. 4 we briefly outline a possible interpretation of the results in
terms of radiation theories.
In Sect. 5 radiative ages are compared with those derived from dynamical arguments and 
we discuss how the two estimates can be made mutually consistent. 
We also compare the results based on our sample with those obtained for more powerful 
radio sources and discuss the differences we find.
Finally, we comment on the possible occurrence of particle re-acceleration.

\section{Source selection and spectral data}

The B2 sample of radio sources, which  
contains about a hundred objects, was obtained from identification of B2 radio
sources with bright elliptical galaxies (see Colla et al. 1975; Fanti et al.
1978).  
Because  of the selection criteria, the sample is dominated by
radio galaxies with a power typically  between $10^{23}$ and $10^{25}$ WHz$^{-1}$
at 1.4~GHz. 
Most are FRI sources. A few have a Head-Tail (HT) or Wide-Angle-Tail (WAT) structure, or
are 3C31-like objects, i.e. sources whose 
brightness fades gradually with distance from the core. The majority, however,
are double sources with bright symmetric jets,  while hot-spots are usually weak or
even absent.
The  sample has been extensively studied with different VLA configurations
at 1.4 GHz (see references in Fanti et al. 1987). More recently
Morganti et al. (1997a) observed 56 sources of the sample at 5 GHz, 25 with D array, 
21 with C array and 10 with B array. 
These observations match the resolution of previous 20 cm observations
and were originally planned in order
to study depolarization asymmetries (Morganti et al. 
1997b) in the lobes at a moderate resolution.

From the above sample we have selected sources which fulfil the 
following  criteria:  

\begin{description}
\item{a)} a ratio of overall source size to beam size $\gtae $ 10;
\item{b)} a signal to noise ratio per map point $>$ 5, at both frequencies,  
for at least 8 independent  slices, separated by at least one beamwidth, 
and perpendicular to the source axis, in order to  
obtain a trend of the spectrum along  the source.
\end{description}

This leaves us with 32 sources of the original sample. Of these, 29 
have double lobed morphology, 2 are WATs and 1 is a NAT.
Due to criterion a), these sources tend to be the ones with  
larger angular and linear size. 
For some objects a spectral analysis had already been done in the past 
(Morganti et al. 1987; 
Capetti et al. 1995).  However, these data have been re-analysed here in
order to ensure homogeneity.

We have carefully checked if the maps account for the total flux
density (as measured 
by lower resolution observations), in order to avoid spurious spectral trends 
due to missing flux at one of the two frequencies or even at both. The VLA fluxes 
and those at lower resolution generally agree within the combined error, 
except for a few sources which were therefore excluded from the following analysis.

Figure 1 shows how the sources of the B2 subsample (with spectral index 
information\footnote{We use the convention $S\sim \nu^{-\alpha}$}
either presented here or available in the literature; see below) are distributed in the radio 
power---linear size plane, and how this compares with the complete B2 sample.
Clearly, for linear sizes larger than 10 kpc the sub-sample 
can be considered to be representative of the complete B2 sample.

For each source we have found the average of the two frequency spectral index 
$\alpha_{5.0}^{1.4}$ 
for slices perpendicular to the source axis and thus 
produced the variation of the spectral index along the source major axis.
The slices are one beam across, so that the data points are practically 
independent.
By visual inspection of the intensity maps we decided which
source regions containing radio jets and hot spots were to be excluded.

Spectral index errors along the profiles are largely 
determined by the map noise and
in a minority of cases by dynamic range. They are typically $\approx$ 0.05, but
can be as high as $\geq $ 0.1 in the faintest regions considered.
Unfortunately the cut-off imposed on brightness at 5 GHz introduces a bias
against the inclusion of regions with very steep spectra. Possible cases of such 
steep-spectrum emission are noted in Sect. 3.

In most objects the spectral index clearly varies 
with distance from the core. 
The spectrum either steepens from the lobe outer edge inward (hereafter
referred as ``spectral type 2''; see 0908+37 in Fig. 2),
or from the core outward (``spectral type 1''; see 1621+38 in Fig. 2). 
Only a minority of sources does not 
show any significant spectral index trend (these are given as 
spectral type 3 in Table 1).
In Fig.~3 we show the behaviour of the spectral index separately for 
type 1 and type 2 sources.
The ultra-steep source B2~1626+39 is the only one
not used in Fig.~3.

\begin{table*}
\caption[]{Source parameters} 
\begin{flushleft}
\begin{tabular}{clrcrrrllcll}
\hline\noalign{\smallskip}
  B2  & $\alpha_{max}$ & $\nu_{\rm br-min}$ & $\alpha_{inj}$ & B$_{eq}$ & Age  
& LS  & LS$_{\rm n}$ & $\log P_{1.4}$ & spectral & FR & \\
 name       &         &    GHz         &    & $\mu$G   & Myrs & kpc & 
             & WHz$^{-1}$     & type    & type & \\ 
\noalign{\smallskip}
\hline\noalign{\smallskip}
0034+25 &   1.3  & 4.4     & 0.63 &  2.5 &  67   &  83 & 3.30 & 23.13 & 2 & 
I WAT & \\
0206+35 &   1.2  & 6.2     & 0.63 &  5.6 &  35   &  47 & 0.42 & 24.50 & 2 & 
I & \\
0755+37 &   0.75 & $>$30.0 & 0.71 &  4.4 & $<$19 &  79 & 0.53 & 24.49 & 3 & 
I & \\
0828+32 &   0.7  & $>$25.0 & 0.48 &  2.3 & $<$27 & 220 & 1.54 & 24.71 & 2 & 
II & \\
0836+29 &   1.2  & 15.1    & 0.77 &  3.6 &  28   & 351 & 2.45 & 24.73 & ? & 
I-II & \\
0844+31 &   1.4  & 10.0    & 0.84 &  2.4 &  40   & 266 & 1.66 & 24.80 & 1 & 
I-II & \\
0908+37 &   1.1  & 6.0     & 0.66 &  7.4 &  24   &  66 & 0.39 & 24.84 & 2 & 
I-II & \\
0922+36 &   1.2  & 5.4     & 0.45 &  5.0 &  36   & 280 & 1.40 & 24.94 & 1 & 
I-II & \\   
1005+28 &   0.95 & 2.8     & 0.79 &  2.0 &  58   & 403 & 4.28 & 24.25 & 2 & 
I-II & \\
1102+30 &   1.05 & 6.3     & 0.63 &  2.5 &  49   & 160 & 1.84 & 24.27 & 2 & 
I & \\
1113+29 &   0.95 & 21.0    & 0.73 &  6.3 &  17   &  61 & 0.45 & 24.67 & 2 & 
I-II & \\
1116+28 &   1.3  & 4.3     & 0.60 &  2.2 &  62   & 229 & 2.35 & 24.39 & 1 & 
I WAT & \\
1254+27 &   1.4  & 7.7     & 0.66 &  5.0 &  35   &  15 & 0.94 & 22.63 & 1 & 
I & \\
1322+36 &   0.75 & $>$30.0 & 0.64 & 16.0 &  $<$4 &  14 & 0.41 & 23.42 & 3 & 
I & \\
1347+28 &   0.8  & 11.6    & 0.56 &  7.3 &  18   &  48 & 0.71 & 24.05 & 2 & 
I-II & \\
1357+28 &   0.85 & $>$30.0 & 0.80 &  3.3 & $<$22 & 116 & 1.76 & 24.03 & 3 & 
I & \\
1422+26 &   0.9  & 14.4    & 0.66 &  4.5 &  28   &  72 & 1.03 & 24.00 & 2 & 
I & \\
1441+26 &   1.05 & 3.0     & 0.72 &  2.3 &  75   & 190 & 2.88 & 23.97 & 2 & 
I-II & \\
1455+28 &   0.8  & $>$30.0 & 0.79 &  5.6 & $<$13 & 348 & 1.74 & 25.22 & 3 & 
II & \\
1521+28 &   1.0  & 17.5    & 0.72 &  3.0 &  28   & 210 & 1.52 & 24.58 & 1 & 
I & \\
1525+29 &   1.0  & 18.3    & 0.78 &  9.0 &  12   &  20 & 0.30 & 23.98 & 2 & 
I & \\
1528+29 &   1.2  & 17.7    & 0.87 &  2.2 &  29   & 254 & 3.76 & 24.21 & 1 & 
I-II & \\
1609+31 &   0.65 & 21.2    & 0.54 &  9.6 &  10   &  30 & 0.39 & 24.14 & 2 & 
I-II & \\
1613+27&   0.75 & $>$30.0 & 0.67 &  9.0 &  $<$9 &  27 & 0.26 & 24.03 & 3 & 
I & \\
1615+32 &   1.6  & 3.0     & 0.62 &  5.7 &  41   & 160 & 0.76 & 25.79 & 2 & 
II & 3C 332\\
1621+38 &   0.9  & 14.6    & 0.59 &  9.8 &  12   &  17 & 0.39 & 23.65 & 1 & 
I NAT & \\
1626+39 &   2.1  & 0.56    & 0.59 & 10.4 &  57\rlap{$_{\rm KP}$} & 44 & 0.37 & 
24.49 & 1 & I & 3C 338\\
1643+27 &   0.8  & 16.7    & 0.52 &  5.2 &  20   & 180 & 2.55 & 24.05 & 1 & 
I-II & \\
1658+30 &   0.9  & 8.4     & 0.59 &  2.8 &  46   &  78 & 1.42 & 23.88 & 2 & 
I-II & \\
1726+31 &   1.3  & 6.5     & 0.67 &  7.0 &  23   & 197 & 0.99 & 25.89 & 2 & 
II & 3C 357\\
1827+32 &   1.3  & 6.4     & 0.48 &  1.8 &  50   & 312 & 4.00 & 24.07 & 1 & 
I & \\
2236+35 &   0.9  & 11.7    & 0.61 &  8.0 &  17   &  20 & 0.56 & 23.47 & ? & 
I & \\
\noalign{\smallskip}
\hline
\end{tabular}
\end{flushleft}
\smallskip
Column 1 : Source name in the B2 catalogue, with reference to 1950 coodinates;
2: larger observed spectral index; 3: minimum break frequency, from the fit;
4: assumed injection spectral index;
5: equipartition magnetic field;
6: source age, from the JP model (except for 1626+39);
7: source linear size;
8: source size, normalized to the median value for its radio luminosity;
9: radio luminosity;
10: spectral class (see text);
11: Fanaroff-Riley type, taken from Morganti et al. (1997b);
12: 3C name.
\end{table*}

\begin{figure}
\vspace{8.0cm}
\includegraphics{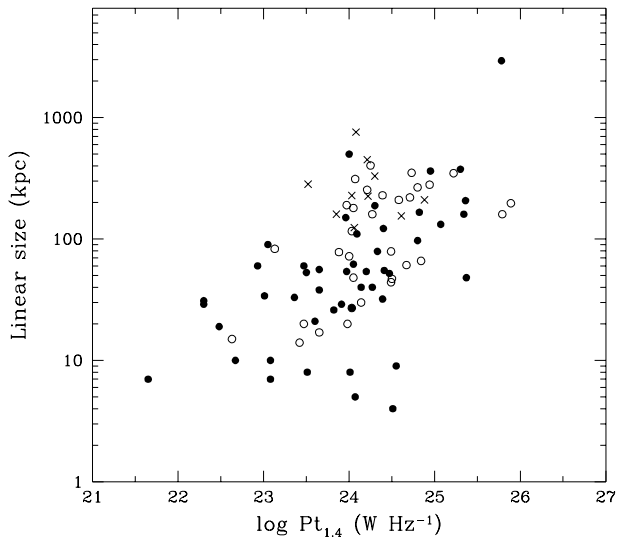}
\caption[]{Linear size as a function of radio power (at 1.4 GHz).
 Open circles: sources with spectral index calculated
from VLA data; crosses: sources with spectral index taken from the literature;
filled circles: B2 sources for which no spectral index information  is
available.}
\end{figure}
 
\begin{figure}
\vspace{5.0cm}
\includegraphics{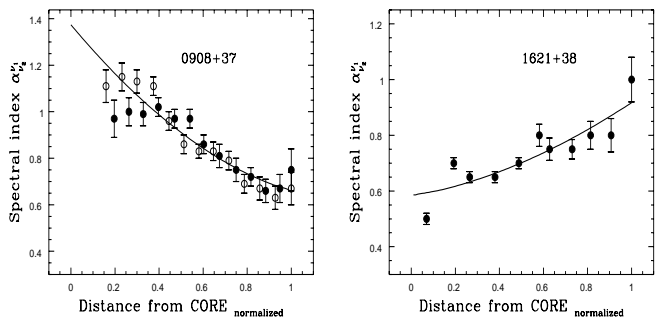}
\caption[]{Two examples of the spectral index distribution as a function of distance
from the core. The distances are normalized to the maximum extent of a lobe.
0908+37 is a double lobed source and the filled and open circles refer to 
the two lobes. 1621+38 is a NAT  source. The full lines are 
the best fit of the radiative model described in the text.}  
\end{figure}

\begin{figure}
\vspace{8.0cm}
\includegraphics{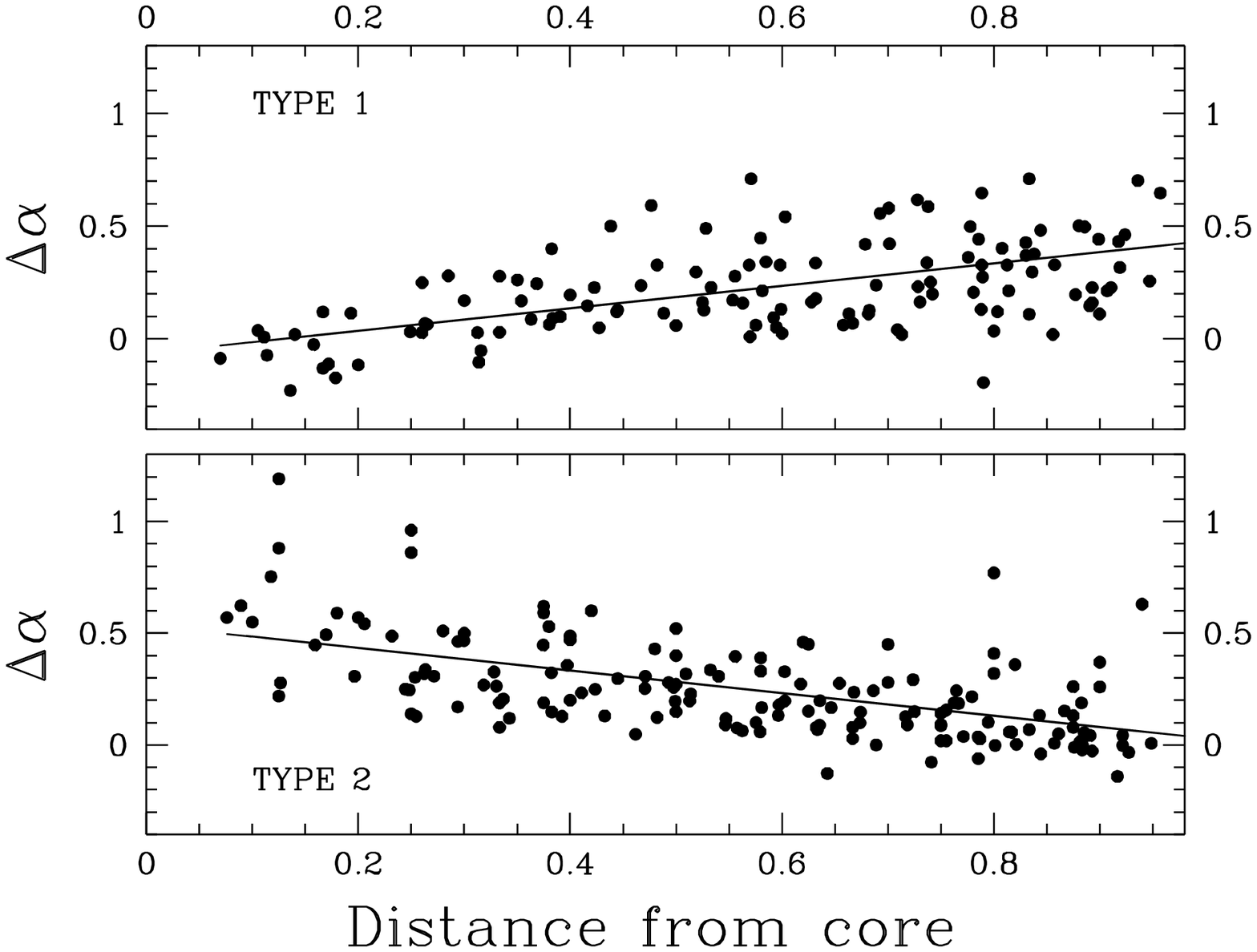}
\caption[]{$\Delta\alpha = \alpha - \alpha_{o}$ as a function of
distance from the core, for type 1 and type 2 sources. For the definitions of $\alpha_{o}$ and spectral type
see the text. The distance
from the core is normalized to the lobe extent. The data
of all sources except B2~1626+39 were used.}
\end{figure}

From a literature search we were able to recover additional information 
on the spectral
index distribution, for 15 objects in the B2 radio galaxy sample.
Their spectral data generally come from frequency pairs other than 
1.4/5 GHz. We have re-analyzed these data with criteria similar to ours
and using the same ageing model (JP, see Sect. 4), in order to ensure 
as much as possible homogeneity. The data are reported in Table 2 and 
individual notes are given in the next section. The meaning of the columns 
is as for Table 1.

\begin{table*}
\caption[]{Sources taken from the literature} 
\begin{flushleft}
\begin{tabular}{clrlrrrlll}
\hline\noalign{\smallskip}
  B2 & Freq. & Age & $\log P_{1.4}$   & LS  & LS$_{\rm n}$ &Spectral &Ref. & FR \\
 Name& range & Myrs & WHz$^{-1}$  & kpc & &type& & type \\ 
\noalign{\smallskip}
\hline\noalign{\smallskip}
0055+30  & 0.3/0.6 &   112 & 24.08 &  760 & 17.9 & 1& 1 & I & NGC315\\
         & 0.6/1.4 &       &       &      &      &  & 9 &   & NGC326\\
0055+26  & 1.4/5   &    35 & 24.61 &  155 &  1.2 & 2& 2 & I \\
0104+31  & 2.7/5   &    93 & 24.21 &  450 &  4.9 & 1& 3 & I & 3C 31\\
         & 0.6/1.4 &       &       &      &      &  & 9 &   \\
0326+39  & 0.6/1.4 &    64 & 24.06 &  124 &  2.1 & 2& 4 & I \\
         & 0.6/11  &       &       &      &      &  &6  &   \\
0828+32  & 0.6/1.4 &    59 & 24.71 &  220 &  1.5 & 2& 5 & II \\
         & 0.6/11  &       &       &      &      &  &6  &    \\
0836+29  & 0.6/1.4 &    30 & 24.73 &  351 &  2.4 & 1& 6 & I-II \\
0844+31  & 0.6/1.4  & $<$35 & 24.80 &  266 &  1.6 &1& 9 & I-II \\
0924+30  & 0.6/1.4 &    69 & 23.52 &  283 &  7.3 & 2&10 & II\\
         & 0.6/11  &       &       &      &      &  &11 &   \\
1243+26  & 0.6/11  &    39 & 24.23 &  129 &  2.3 & 1& 6 & I \\
1321+31  & 0.6/11  &    35 & 23.85 &  160 &  3.4 & 2& 6 & I \\
1358+305 & 0.3/1.4 &    62 & 25.27 & 1330 &  9.1 & 2& 7 & II \\
1615+35  & 0.6/1.4 &    78 & 24.30 &  330 &  3.7 & 1& 8 & I \\
         & 0.6/11  &       &       &      &      &  & 6 &   \\
1827+32  & 0.6/11  & $>$30 & 24.07 &  312 &  4.0 & 1& 6 & I \\
2229+39  & 0.4/2.7 &    74 & 24.03 &  228 &  3.1 & 1& 3 & I & 3C 449\\
         & 2.7/5   &       &       &      &      &  &3  &   \\
         & 0.6/1.4 &       &       &      &      &  &9  &   \\
2335+26  & 0.6/1.4 &    60 & 24.88 &  210 &  1.2 & 1& 9 & I & 3C 465\\
\noalign{\smallskip}
\hline
\noalign{\smallskip}
\end{tabular}

References: 1) Mack et al. (1998); 2) Ekers et al. (1978a); 3) Andernach et al. (1992);
4) Bridle et al. (1991); 5) Parma et al. (1985); 6) Klein et al. (1995);
7) Parma et al. (1996); 8) Ekers et al. (1978b); 9) J{\"a}gers (1981); 
10) Ekers et al., 1981; 11) Klein et al., 1996
\end{flushleft}
\end{table*}
Four of these sources also belong to our VLA sample; 
in Sect. 3 we comment on the spectral ages derived from the different 
sets of data. 

\section{Notes on individual sources}

$\bf 0034+25$. The source is a WAT. The spectral index is well sampled in 
the eastern arm only, which is straight.
The south-western arm, which is strongly bent, 
has a lower brightness, so that 
the spectral index can be measured only in a few 
areas close to the outer edge.

\smallskip\noindent
$\bf 0055+30$. This is a well studied radio galaxy. However the spectral
properties of this source are still contradictory. J{\"a}gers (1981) gives a
spectral index profile along the source (between 0.6 and 1.4 GHz), which shows
a clear steepening outwards, but also large fluctuations are
visible. The average trend would indicate a minimum break frequency
$\geq 1.6$ GHz and a maximum source age $\approx 10^2$ Myrs. 
Mack et al.(1998), based on multifrequency data including Effelsberg 10.5 GHz , 
compute the spectrum in various areas and find large fluctuations
from region to region. They estimate ages $\leq 12$ Myrs. 
We favour the longer timescale, although we feel that the situation is 
rather uncertain.

\smallskip\noindent
$\bf 0828+32$. The radio source has a cross shaped structure
with two prominent low brightness 
wings almost perpendicular to the much brighter lobes 
(Parma et al. 1985). 
The source has been discussed in terms of precession (see also Klein et 
al. 1995). Our VLA data allow us to  
study the brighter lobes only, which
do not show any clear steepening (consistent with what was found by
Parma et al. 1985). The wings have much steeper spectra
(Parma et al. 1985;  Klein et al. 1995). The source age in Table 1
refers to the bright lobes, while the one in Table 2 is obtained from the 
spectral steepening in the wings.

\smallskip\noindent
$\bf0836+29$. The source shows a clear spectral steepening in the western tail
of the southern lobe, which begins at the hot spot position and is almost
perpendicular to the source main axis.
The spectral index may go up to 1.5 (as considered in Capetti et al. 1995), 
but
in the present analysis we have considered only areas where the signal to noise
ratio is $>$ 5: we excluded therefore the steeper spectrum areas. 
The northern lobe is so small that it is impossible to follow any spectral trend.
Besides our 1.4/5 GHz VLA data (Table 2), the source spectrum has been studied
with different resolution in the frequency range 0.6/10.5 GHz (Table 2). The
source ages determined from the two different frequency ranges are in 
reasonable agreement. 

\smallskip\noindent
$\bf 0844+31$. The source was studied by Capetti et al. (1995). 
Both lobes contain hot spots. The spectrum steepens from the hot spots outward.
In the southern lobe, the spectral index appears to reach a saturation value,
after an initial steepening. The same may happen in the northern lobe. This behaviour 
is discussed in terms of re-acceleration processes in Sect. 5.4.

\smallskip\noindent
$\bf 0922+30$. The spectrum of the southern lobe steepens outwards from 
a hot spot, which is located well inside the lobe.
The northern lobe is too small to be able to see any spectral trend.

\smallskip\noindent
$\bf 0924+30$. See also Cordey (1987), for a discussion of this source, 
which is believed to be a remnant.

\smallskip\noindent
$\bf 1005+28$. The spectrum steepens inward in the lobes. However, due 
to the low brightness of the source, we can follow the spectral steepening 
only for 
about half the length of the lobes. The estimated age is derived by 
extrapolating the observed spectral trend back to the core.

\smallskip\noindent
$\bf 1116+28$. The source is a WAT. The radio spectrum steepens away 
from the radio core
and seems to saturate at about 2/3 of the arms' length.
This behaviour 
is discussed in terms of re-acceleration processes in Sect. 5.4.

\smallskip\noindent
$\bf 1441+26$. See note for 1005+28.

\smallskip\noindent
$\bf 1521+28$. See Capetti et al. (1995). The southern lobe is very long 
and contains 
a weak hot spot in the middle, from where the spectral index increases 
outwards and then saturates at a constant value. This behaviour 
is discussed in terms of re-acceleration processes in Sect. 5.4.
No spectral trend can be followed in the northern lobe, which is too small.

\smallskip\noindent
$\bf 1528+29$. The source has a double lobed morphology with rather bright
twin jets.
The spectrum steepens outwards and seems to saturate at about half lobe 
length. This behaviour 
is discussed in terms of re-acceleration processes in Sect. 5.4.

\smallskip\noindent
$\bf 1615+35$. This is a NAT source. Ekers et al. (1978b) show that the 
spectral
index, after an initial steepening, saturates at $\approx$ 1.1.

\smallskip\noindent
$\bf 1621+38$. The source is a small NAT. 

\smallskip\noindent
$\bf 1626+39$. The radio spectrum is very steep all over the source, 
perhaps slightly steeper in the outer regions. The JP model is unable to 
fit the spectral index distribution, while the KP model is rather satisfactory.
The age in Table 1 is from this last model.
An alternative and physically more acceptable fit is discussed in Sect. 5.4.
The source is associated with the brightest galaxy of the cluster A 2199.

\smallskip\noindent
$\bf 2236+35$. The source shows two faint wings, antisymmetric with respect 
to the main 
bright lobes. The spectrum in the bright lobes is almost constant 
($\alpha_{5.0} ^{1.4} \approx 0.6$). However a clear spectral steepening is 
seen in the eastern wing. The source age is derived from this steepening.

\section{Determination of the radiative ages} 

\subsection{The models and the assumptions}

The spectral trends along the sources are interpreted  in 
terms of radiative ageing of the relativistic electrons by synchrotron and
I.C. processes. The analysis of the spectral data follows  well known 
arguments (see, e.g. Myers \& Spangler 1985). 
By assuming an injection 
spectral index, $\alpha_{inj}$, it is possible to obtain  the 
break frequency, $\nu_{br}$, as a function of position along the source 
axis and, finally, for an assumed magnetic field, the
age of the relativistic electrons in that position. 

Theoretical synchrotron-loss spectra have 
been computed numerically (Murgia 1996) from the  synchrotron 
formulae (e.g., Pacholczyk 1970). 
It is assumed that the synchrotron and 
I.C. losses dominate and that expansion losses and re-acceleration 
processes can be neglected (see e.g. Alexander 1987, for a discussion 
of synchrotron losses versus adiabatic expansion).

Two models are generally considered:  i) the Jaffe-Perola
(JP) model (Jaffe \& Perola 1974), in which the time scale for continuous 
isotropization of the 
electrons is assumed to be much shorter than the radiative time-scale;  
ii) the Kardashev-Pacholczyk (KP) model (Kardashev 1962) in which each 
electron maintains its original pitch angle.
We have preferred the JP model (except for 1626+39, see note) since the 
I.C. energy losses due to 
the microwave background radiation are as important
as the synchrotron losses, and in the former random
orientations are expected  between electrons and photons.

The two frequency spectral index $\alpha ^{\nu_1}_{\nu_2}$ allows us
to compute the break frequency, $\nu_{br}$, in various regions of the source, 
using the synchrotron-loss spectrum for the JP  model.
The break frequency is used to determine a spectral age, based on the
formula:
$$t_s = 1.61 \times 10^3 \frac {B^{0.5}}{(B^2+B_{CMB}^2)(\nu_{br} (1+z))^{0.5}}$$
where the synchrotron age $t_s$ is in Myrs, the magnetic field B is in 
$\mu G$,
the break frequency $\nu_{br}$ in GHz and $B_{CMB} = 3.2 \times (1+z)^2$
$\mu G$ is the equivalent magnetic field of the cosmic microwave background
radiation.
It is assumed that the magnetic field strength is uniform across the source 
and has remained constant over the source life.

The magnetic field is computed using the ``minimum energy assumption''.  
We assumed equality in the energy of protons and electrons, a filling factor 
of unity, a radio spectrum ranging from 10 MHz to 100 GHz and an ellipsoidal 
geometry.
For most of the objects in our sample the computed magnetic field is within a factor 2 of 
$B_{CMB}$. This 
ensures that the synchrotron ages $t_s$ are relatively independent of moderate
deviations from the ``minimum energy conditions''.

Figure 4 shows how the radiative life time depends on the ratio 
$B/B_{eq}$. For $ 0.5 < B_{eq}/B_{CMB}
< 2$, deviations from equipartition have a small effect on the computed 
lifetime if $B \ltae 2 \cdot B_{eq}$.

\begin{figure}
\vspace{8.5cm}
\includegraphics{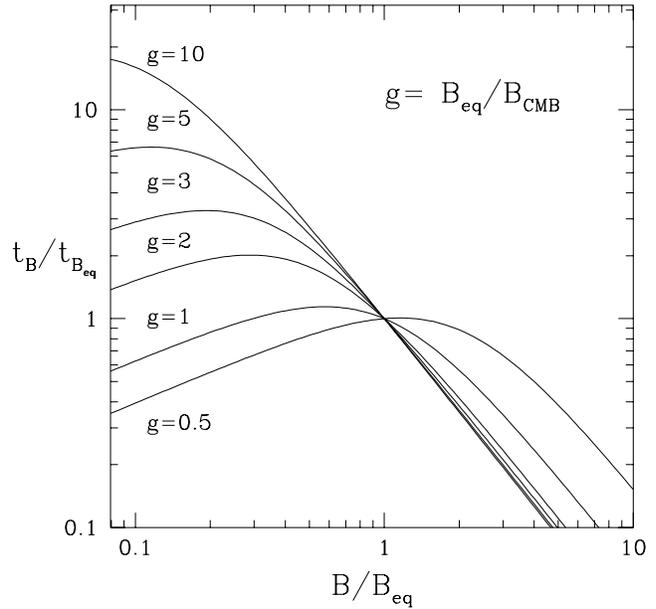}
\caption[]{The radiative lifetime as a function of the ratio $B/B_{eq}$.}  
\end{figure}

\subsection{ Spectral analysis}

For the majority of the sources the quality of spectral information is comparable 
in the two lobes, and the spectral trends are rather similar when the
errors are taken into account. Therefore we preferred to analyse the two lobes
together, instead of considering them separately. 
We did this by folding the spectral profiles of the two lobes 
onto each other, after having normalized the coordinates along the lobe axis to the 
maximum lobe extent (as shown  in Fig. 2, for 0908+37).
We fitted the folded spectral index plots along the source axis 
(as in Sect. 2) with a relationship of the type:
$$\nu_{br} \propto x^{-2}$$
\noindent
where $x$ is the normalized distance from the outer lobe edge or from the 
core according
to the observed steepening trend (as shown in Fig. 2). 
The above law is expected on the
basis of the relation between radiative age and break frequency, if $x$ is
proportional to time, i.e. assuming a constant expansion speed. From it 
we obtain a best estimate of $\alpha_o$, i.e. the spectral  
index close to the outer edge of the lobe or close to the core, according to the
observed spectral profile. This, with a few exceptions discussed below, is
assumed to be equal to the injection spectral index, $\alpha_{inj}$.
We find a mean value of $\alpha_o \approx 0.65$,
with a dispersion $\approx 0.1$.
In this way we obtain from the folded spectral profiles the variation 
of the break frequency as a function of position, averaged 
over the two lobes.
Extrapolating it to the inner or outer part of the source, we then obtain the
``minimum break frequency'', $\nu_{br-min}$, namely the break frequency of the
oldest electrons, that we use for determination 
of the source age.
For the large majority of the sources the extrapolation introduces only minor 
additional uncertainties. 
The computed values of $\nu_{br-min}$  range from a few GHz up to several
tens of GHz.

Considering the errors in the spectral index ($\gtae 0.05$) and the 
uncertainty in $\alpha_{inj}$ ($\ltae 0.1$), the uncertainty 
in $\nu_{br-min}$ can be quantified as:
$$\Delta \nu_{br}/\nu_{br} \ltae 0.08 \cdot (\nu_{br}(GHz))^{0.5}.$$
This implies that only values of $\nu_{br-min}$ $\leq$ 30 GHz 
are significant, which is hardly surprising given that our highest frequency 
is a factor six lower.
The errors on $t_s$ , for a given  magnetic field,
are $\approx$ 15 \%, for break frequencies of a few GHz, increasing up  
to $\geq$ 40 \% for break frequencies $\approx$ 30 GHz.
Had we used the KP model, $t_s$ would be shorter 
by $\approx$ 10 $\%$.

As our analysis of the folded spectra might have missed some information 
on possible differences between the two lobes, we have further 
compared for each source the best fit model with the two lobes separately, 
in order to see how well each of them individually fitted the model. Differences between 
the two lobes appear to be insignificant. The $\alpha_o$ values of the two lobes are 
in general within 0.1 and the values of $t_s$  are within the estimated 
uncertainties.

We have also performed a fit with a relationship $\nu_{br} \propto x^{-m}$,
as a check of the assumption of constant expansion speed. Within the 
uncertainties, we find that $m$ is close to two. We do not claim that
the expansion speed is constant, but only state that our data do not show 
strong deviations from it.

We comment a bit more on $\alpha_{inj}$, since $\nu_{br-min}$ strongly
depends on it. Alexander \& Leahy (1987) and Leahy et al. (1989), 
in their spectral study of several tens of 3CR 
radio sources, have used as $\alpha_{inj}$ the low frequency spectral index,
$\alpha_{low}$, of the integrated spectrum.
For our sources this choice is less appropriate, since  
$\alpha_{low}$ (known between 0.4 and 1.4 GHz) has 
errors $\geq 0.1$: therefore we prefer to use $\alpha_o$. 
Nevertheless we have compared the $\alpha_o$ 
with the $\alpha_{low}$ in order to check consistency. 
For most sources $\alpha_o$ and $\alpha_{low}$ are the same within the expected errors.
However, for sources which have $\nu_{br-min}$ $<$ 10 GHz, $\alpha_{low}$ 
tends to be systematically larger than $\alpha_o$, by $\approx 0.1$
or more. It appears that the integrated spectral index $\alpha_{low}$, even
when measured at low frequencies,
has already suffered somewhat from radiative losses
and therefore it is better to assume that  $\alpha_{inj}$ = $\alpha_o$.
For two sources (0034+25, 0206+35) $\alpha_o$ $\approx 0.9 - 1.0$,
significantly different from $\alpha_{low}$ ($\approx$ 0.6).
For them we assume that $\alpha_o$ is modified from $\alpha_{inj}$ by a high 
frequency break, perhaps due to energy losses in the jet, and assume 
$\alpha_{inj}$ = $\alpha_{low}$.

The radiative lifetimes are in the range of $10^7 - 10^8$ years.
We stress that we cannot exclude the possibility that we  have 
missed source areas with steeper spectra. Therefore the $t_s$ we give are
lower limits on the source ages. 

\section{Discussion}

\subsection{Two classes of source}

There appear to be two classes of source in our sample (spectral types 1 
and 2), according to the run of the $\alpha ^{1.4}_{5.0}$ along 
the source axis.
This double behaviour was already pointed out some time ago by J{\"a}gers (1981).
In the  ``spectral type 2'' sources,  $\alpha ^{1.4}_{5.0}$ increases 
from the outer edges of the lobes towards the core.
Therefore the older electrons are found closer to the core. 
This class contains essentially  
sources with double lobed morphology, of both FR~I and FR~II type.
Their spectral behaviour is that 
expected on the basis of standard source models where the radiating particles 
are deposited at different times at the end of an advancing beam and remain
in that position or flow back at some speed toward the core. In this model 
the closer to the core the older they are. 

In the ``spectral type 1'' the spectral index steepens away from the core.
This class is known to contain ``3C 31 like'' objects, WATs, and  NATs 
(see, e.g., J{\"a}gers, 1981). In our sample of Table 1 the
spectral class 1 objects are: one WAT, one NAT and some double
sources with plumes or wings where this spectral behaviour is seen. 
In this last type of objects hot spots, if present, are seen well inside the 
lobes and the steepening of the spectrum starts from there. 
One double source with very bright twin jets and narrow lobes (1528+29) also 
belongs to this class.

The distributions of radiative ages, $t_s$, for the two spectral classes 
are very similar.

\subsection{A comparison between spectral and dynamical ages}

We have compared the synchrotron ages, $t_s$, with the dynamical ages, $t_d$,
evaluated from simple ram-pressure arguments.
The expansion velocity of the lobes is given by the relation:
$$v_{exp} \approx \left[\frac{\Pi}{A}~ \frac{1}{m_p  n_e}\right]^{0.5}$$
where $\Pi$ is the jet thrust and $A$ is the size of the area over which 
it is discharged. 

The most simplistic assumption is to identify the $\Pi/A$ ratio 
with the hot spot pressure. However various authors have suggested that the 
thrust is likely to be applied over a wider area and therefore that the 
deduced velocities are overstimated (see, e.g.: Norman, 1993 and Massaglia 
et al., 1996, where differences of up to a factor 3 are shown).
The most popular mechanism for accomplishing this is the Scheuer's  ``dentist
drill'' scenario (Scheuer 1982), according to which the jet direction fluctuates 
on short timescale compared to the age of the source. In order to reduce this 
problem,  we use, the front surface minimum pressures of the lobes, 
$p_{eq,f}$, as the appropriate quantity 
for $\Pi/A$ (Williams, quoted by Carilli et al., 1991).


The values we  
find are typically $\leq 4$  times the average minimum lobe pressure. 


We assume that the jet thrust is constant over the source 
life time and that the source grows in a self-similar way, such that  
$p_{eq,f}  \propto R_{kpc}^{-2}$. 
Finally we assume a run of the external density $n_e = n_o \times 
R_{kpc}^{-\beta}$, with $n_o \approx 0.5~$ cm$^{-3}$
and $1.5 \ltae \beta \ltae 2$,  according to Canizares et al. (1987).
Under these assumptions there would be only  slight dependence, if any, of 
$v_{exp}$ on source size, depending on the value of $\beta$.

We have excluded from this analysis the NAT source 1621+38, for which the
ram-pressure model is likely to be incorrect.  We have, instead, included
the two WATs, although this may be questioned, since the jets are moderately 
bent in 1116+28 and one of them is straight in 0034+25, and they run 
undisrupted to the end of two lobes. In any case the exclusion of these 
two sources does not change our conclusions.

In Fig. 5 we show the plot  $t_s$  $\it vs$  $t_d$ , for $\beta = 2.0$.
The dots represent sources with spectral type 1, 
the open circles referring to four sources which are discussed in Sect. 5.4
and whose radiative lifetimes could be definitely longer than the values 
given in Table 1. Sources with spectra of type 2 are denoted by triangles.
The arrows refer to sources for which the $t_s$ are upper 
limits.

The dynamical and radiative ages are correlated, but the dynamical ages
in general are larger by a factor $\approx 4$ for $\beta = ~1.5$ and 
$\approx ~2$ for $\beta = 2$. 

\begin{figure}
\vspace{9.0cm}
\includegraphics{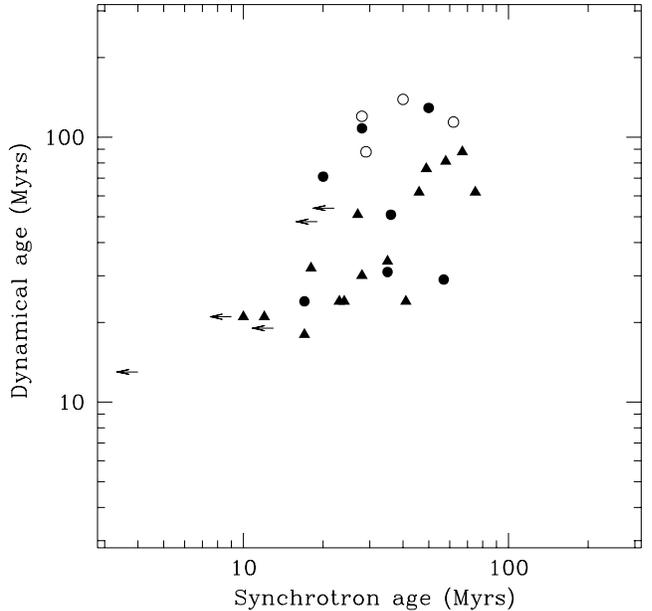}
\caption[]{Dynamical age vs  spectral age of the source. For the meaning of the
symbols see text.}
\end{figure}

These discrepancies are believed to be not very serious for at least three
reasons. 
First of all, the  one-dimensional ram-pressure balance may not be 
realistic and discrepancies of a factor 2 or so are well possible, as 
mentioned above.
Second, the central density we have assumed may be a bit too high.
A value around 0.1 cm $^{-3}$, which is not excluded by the X-ray data, 
would bring, for $\beta~ = ~2$, the dynamical ages into closer agreement with
the spectral ones. Third, the ram-pressure dynamical 
ages depend on the minimum energy assumption. It has been shown
that radio galaxies with luminosities like in the present sample usually
have internal pressures larger than the minimum ones by a factor $\geq 5$
(Morganti et al. 1988; Feretti et al. 1992).
If we assume that the magnetic field is weaker than the one corresponding
to minimum energy conditions by a factor $\geq$ 4, the internal 
pressure increases 
by a factor $\geq$ 4 and the dynamical lifetime decreases by a factor 
$\approx$ 2.
Anyway, the radiative lifetimes would change little, as can be 
seen from Fig. 4.
Likely  both explanations may play a role in bringing $t_s$ and $t_d$ to
a closer agreement.

\subsection{Spectral ages and source sizes}

We have investigated whether there is a relationship between  radiative age and
source size. 
We find a significant correlation between radiative age and linear size, 
as shown in Fig.~6.
A linear fit between the logarithms of age and linear size gives:
$$LS \propto {t_{s}}^{0.97},$$
\noindent
where the uncertainty in the exponent is 0.17. 

The advance speed of the lobes, deduced from the synchrotron ages, are in the 
range of $0.5 - 5 \cdot 10^3$~km/sec. There is no difference between the 
two spectral classes.

Alexander \& Leahy (1987) have presented a strong correlation between the 
expansion speed, deduced from the radiative lifetimes,  and radio power.  
Liu et 
al. (1992) comment also on this relation and point out that it could be 
caused partly by the assumption of $B = B_ {eq}$. It should be noted that 
their sources are of spectral type 2. The velocities we have 
deduced from our sample, relatively independent of the equipartition 
assumption, are in good agreement with those of the low power sources in 
those samples.

\begin{figure}
\vspace{9.0cm}
\includegraphics{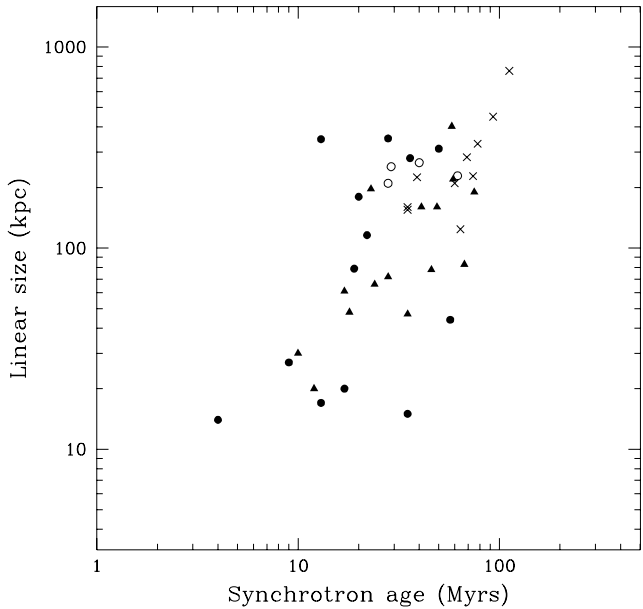}
\caption[]{Linear size as a function of spectral age of the source.
The crosses represent the sources taken from the literature, the dots type 1
spectra, the open
circles, as in Fig. 5, referring to the four sources discussed in Sect. 5.4.
The triangles represent type 2 spectra.}
\end{figure}

\subsection{Is there particle re-acceleration?}

As mentioned in Sect. 4, we have analysed our spectral data
assuming that the dominant processes producing the energy losses of the 
relativistic electrons are synchrotron and inverse Compton, neglecting 
adiabatic expansion as well as re-acceleration. 
With the additional assumption of a constant advance speed of the
source outer edges, this is the motivation for the fit  to the spectral
data, $\nu_{br} \propto x^{-2}$, that we have used in Sect. 4.2. The fits 
to the data in general are good, in the sense that they 
describe well the overall trend of $\alpha_{5.0}^{1.4}$ versus $x$. 

There are however a few objects were the break frequency does not 
seem to increase
always with $x$ as expected from the simple synchrotron/I.C. model. 
An example is shown in Fig. 7.

\begin{figure}
\vspace{8.0cm}
\includegraphics{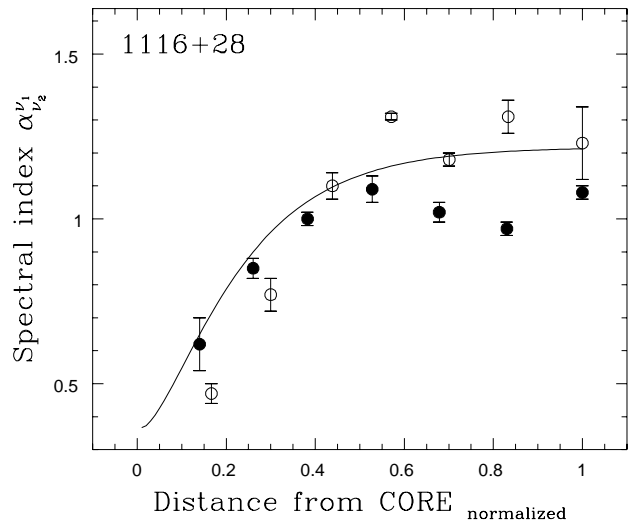}
\caption[]{Spectral index as a function of distance from the core, for the
source B2~1116+28. The  line represent the re-acceleration model described
in the text.}
\end{figure}

The saturation of $\alpha_{5.0}^{1.4}$ to a constant value would 
indicate that the break frequency, after an initial decrease,  
does not decrease anymore. 
A possible explanation of this is that a re-acceleration process is acting,
which compensates in part for the radiative energy losses and causes a freezing 
of the break energy at that value where the radiative and the acceleration 
time scales are equal. Of course, if re-acceleration processes are working, the
ages estimated in the previous section would be  underestimated.
Expressing the re-acceleration process as:

\smallskip
$dE/dt = E/ \tau_a,$
\smallskip

where E is the particle energy, and $\tau_a$ is the  
acceleration time scale, the break frequency, as a function of time
(see Kardashev, 1962), is given by:
$$\nu_{br} \propto \frac{B}{(B^2+B_{CMB}^2)^2} \frac{1}{(1-e^{-t/ \tau _a})^2 ~ 
\tau_a^2}
\propto  \frac {\rho^2}{(1 - e^{- \rho ~x})^2},$$
\noindent
where $\rho$ is the ratio of the source age to the re-acceleration 
time scale $\tau _a$ and x is the normalized distance from the outer lobe 
edge or from the core according to the observed steepening trend.

The above law naturally leads to a saturation of $\nu_{br}$ when 
$t \gtae \tau _a$.

We have re-fitted all the spectral profiles with the above formula and derived 
a value of $\rho$ for each source and corresponding source age, 
$t_s'$, which is related to the previous ones, $t_s$ by:
$$t_s' = \frac{t_s \rho}{1- e^{- \rho}}$$
For at least 70\%  of the objects the results are not much different from the 
purely radiative model. For those we find  $0 \leq \rho \leq 2$ 
(the median value
is 1.1) and the 
lifetimes are modified by less than a factor 2. Time scales for re-acceleration
must be typically $\geq 4 \cdot 10^7$ years.

However for a few  sources (0844+31, 1116+28, 1521+28, 1528+29)  
we find $\rho$ $>$ 5, which suggests  that re-acceleration may be present. 
Their ages would then be raised by factors from $\approx$ 6 to $\approx$ 
10 with respect to the ``radiation only'' model.
These  objects seem to be
among those  where the spectrum steepens away from the core. 
Also 1626+39 is fitted reasonably well by this model and this is physically
more acceptable than the KP model (see note in Sect. 3), but its age is 
only marginally modified.


Finally, we note that the possibility of re-acceleration processes is 
another factor which goes in the direction of bringing the radiative 
and the dynamical timescales closer.

\subsection{Spectral ages and radio power: a comparison with the 3CR 
radio galaxies}

We have compared the synchrotron ages derived for our sample with the 
corresponding ones for 3CR sources found by Alexander \& Leahy (1987);
Leahy et al. (1989) and Liu et al.(1992). When necessary the data were
corrected, to take into account different assumed values for
the Hubble constant.

Figure 8 shows a plot of the ages versus radio power for the B2 and the 
3C samples.
\begin{figure}
\vspace{9.0cm}
\includegraphics{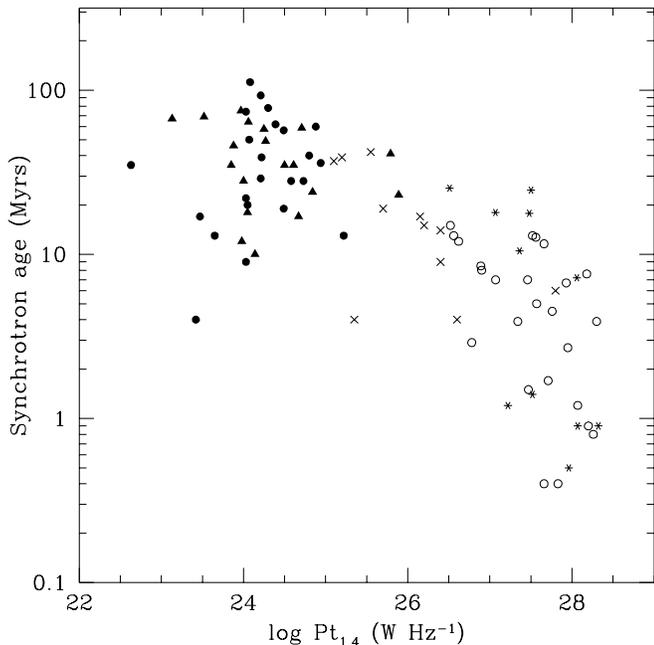}
\caption[]{Source age versus radio power at 1.4 GHz. The meaning of the symbols is
as follows. Filled circles and triangles: B2 sources with type 1 and type 2 spectra, respectively; 
crosses: 3C galaxies with $z<0.2$; open circles:
3C galaxies with $z>0.2$; asterisks: 3C quasars.}
\end{figure}
It appears that the $t_s$ of our sources are systematically larger than those
of the 3CR sources by factors 5 - 10. 
In this context it is relevant to mention that Liu et al. (1992) noted that the synchrotron 
lifetimes of the more 
powerful 3CR sources are shorter than those of low-power sources.
The trend of synchrotron age vs. radio power corresponds to a correlation
between expansion speed and radio power, as recalled in Sect. 5.3,  since the 
sizes of the sources in the two samples are not too dissimilar.
There are several possible reasons for these differences, among which we 
mention the following.

a) As discussed by  Liu et al. (1992), the radiative ages derived for the 
powerful sources (3CR) are  heavily
dependent on the equipartition magnetic field, $B_{eq}$, which is 
generally significantly larger than $B_{CMB}$. 
For B $\approx ~ 1/4 \cdot B_{eq}$ the differences between the two sample 
would be greatly reduced.

b) Another possibility is that, if there is significant backflow in the 
powerful sources, in the inner regions there is an accumulation and
mixing of the older electrons, with a consequent smoothing of the spectral 
break and an under-estimation of the maximum radiative time scale.

c) The effect could be real, namely  in  lower power sources, as those of 
our sample, the nuclear activity lasts for a longer time. 
One could think of an evolutionary effect, in the sense that low power
sources have evolved from high power sources and are therefore older.
We note, however, that the sources of our sample are on average smaller
than powerful 3CR sources (de Ruiter et al. 1990), so that we consider this
unlikely in general, even if it cannot be excluded as an explanation in
some cases.  
Another possibility is a cosmological effect, since the 
3CR sources are at much larger red-shifts than B2 radio galaxies.

We note that Scheuer (1995), based on an analysis of the lobe asymmetries
of powerful radio  sources, concluded that their growth rates are likely to be
less than $0.1 ~c$ and could be as low as $0.03 ~c$. This would indicate 
source lifetimes that are longer than the radiative ones and closer to those we 
find for the B2 radio galaxies.

\section{Conclusions}

$ \it 1)$ We have derived the variations of the spectral index across the source 
emitting
regions for a representative sample of 32 radio galaxies from the B2 sample,
by means of VLA maps at 1.4 and 5 GHz observations. From the literature 
we have found similar information for 
an additional 10  B2 radio galaxies, bringing the total number to 42.
From these data, using 
a simple standard radiative ageing model, we have computed the radiative ages.

\medskip

\noindent
$\it 2)$ The typical ages we find are in the range of $10^7 ~- 10^8 $ years,
somewhat longer than the values found for more powerful 3CR radio sources.
We discuss this point briefly and suggest that moderate deviations from the
equipartition assumption, on which the age determination is based, may 
easily explain the discrepancy. However we cannot exclude that the effect is 
real.

\medskip

\noindent
$\it 3)$ We compare the radiative ages with dynamical ages obtained from a
simple ram-pressure model. We find good statistical agreement between 
the two methods, for an average central gas density of the ambient medium
$\approx 0.1 $~cm$^{-3}$ and for moderate deviations from the equipartition 
conditions.
A correlation is found between the linear size, $LS$, and the
radiative age.

\medskip

\noindent
$\it 4)$ We discuss the possibility of re-acceleration of relativistic
electrons in the source. A few objects appear to indicate the
existence of such an effect. The time scale of re-acceleration is
typically $\geq 4 \cdot 10^7$ years.

\medskip

\noindent
$\it 5)$ Our spectral interpretation is based on the generally assumed ``standard model'', 
namely an aged electron power law in a uniform magnetic
field. In recent works this approch has been criticised (e.g. Katz-Stone et 
al., 1993; Eilek \& Arendt,1996). It is claimed that  high frequency spectral 
steepening is  not necessarily due to synchrotron ageing.
With our data we are not in a position to comment on these ``non  orthodox''
views. Were they correct, the standard discussion of our data, as well as
those made in the past, would be invalidated.
However we feel that the reasonable agreement between the standard radiative
ages and the dynamical ages from a simple ram-pressure model are suggestive
that the standard picture is at least partly correct.

\section {Acknowledgements}
We are indebted to Drs. R. Laing and S. Spangler, who carefully read the manuscript and 
provided useful comments. We also thank the referee Dr. J. Riley, whose constructive
criticism helped to improve the presentation of the paper. 
The National Radio Astronomy Observatory is operated by Associated 
Universities, Inc., under contract with National Science Foundation.

\end{document}